\documentstyle{elsart}
\include{psfig}

\begin{document}
\begin{frontmatter}
\title{Nonequilibrium Josephson current in ballistic multiterminal SNS-junctions}

\author{P. Samuelsson, \AA . Ingerman, V.S. Shumeiko, and G. Wendin} 
\address{Department of Microelectronics and Nanoscience,
  Chalmers University of Technology, S-412 96 G\"oteborg, Sweden}
\begin{abstract}
 We study the nonequilibrium Josephson current in a long
  two-dimensional ballistic SNS-junction with a normal reservoir
  coupled to the normal part of the junction. The current for a given
  superconducting phase difference $\phi$ oscillates as a function of
  voltage applied between the normal reservoir and the SNS-junction.
  The period of the oscillations is $\pi \hbar v_F/L$, with $L$ the
  length of the junction, and the amplitude of the oscillations decays
  as $V^{-3/2}$ for $eV \gg \hbar v_{F}/L$ and zero temperature. The
  critical current $I_c$ shows a similar oscillating, decaying
  behavior as a function of voltage, changing sign every oscillation.
  Normal specular or diffusive scattering at the NS-interfaces does
  not qualitatively change the picture.
\end{abstract}
\end{frontmatter}
%

pacs[74.50.+r, 74.20.Fg, 74.80.Fp]

\section{Introduction}
In recent years there has been an increased interest in the
nonequilibrium Josephson current in mesoscopic multiterminal
SNS-junctions. Quasiparticle injection from one or several normal
reservoirs coupled to the SNS-junction leads to a nonequilibrium
population of the current carrying Andreev levels, and thus to a
modification of the Josephson current. As predicted by theory,
\cite{vanwees91,volkov95,chang97,samuelsson97,yip98,wilhelm98}
experiments show that suppression \cite{morpurgo98,schapers98},
switching \cite{baselmans99} and even enhancement \cite{kutchinsky99}
of the Josephson current under nonequilibrium is possible.

The theory for the nonequilibrium Josephson effect has been developed
for mainly two types of junctions, quantum ballistic
\cite{vanwees91,chang97,samuelsson97} and diffusive
\cite{volkov95,wilhelm98,yip98}. A growing experimental interest is
however shown for multiterminal SNS-junctions where the normal part is
a ballistic semiconductor 2DEG.\cite{schapers98,comment1} In this
paper, a theory for these type of structures is presented, and it is
applied to junctions with and without normal reflection at the NS
interfaces. For junctions with normal reflection, both specular and
diffusive scattering is taken into account.
\section{Model and theory}
\begin{figure}[h]
\centerline{\psfig{figure=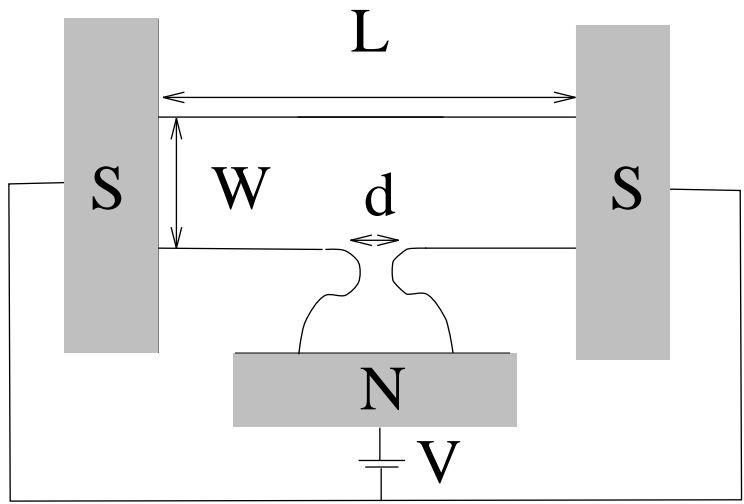,width=0.7\linewidth}}
\caption{A schematic picture of the junction.}
\label{junction}
\end{figure}

A model of the junction is presented in Fig. \ref{junction}. A
ballistic two-dimensional normal region, width $W$, is connected to two
superconducting electrodes, with electrode separation $L$. The phase
difference between the superconductors is $\phi$. A normal electron
reservoir is connected to the normal part of the junction via a
quantum point contact, with width $d\ll W$, and a voltage $V$ is
applied between the normal reservoir and the SNS-junction.

The resistance of the point contact is assumed to be the dominating
resistance of the junction, and the applied voltage thus drops
completely over the injection point. Zero magnetic field is assumed.

A natural framework for studying multiterminal ballsitic
two-dimensional SNS-junctions is the Landauer-B{\"u}ttiker scattering
approach. The junction is described by the Bogoliubov-de
Gennes-equation (BdG), where it is assumed that the superconducting
gap $\Delta$ is constant in the superconductors and zero in the normal
metal. Hard wall boundary conditions for the 2DEG are also assumed. Solutions
to the BdG-equation are matched at the NS-interfaces and at the three
lead connection. It is assumed that all transverse modes couple
equally to the modes in the injection point contact.\cite{schapcom}

Boundary conditions are incoming electron and hole quasiparticles from
the normal reservoir and incoming electron- and hole-like
quasiparticles from the superconductors at energies above the gap.
Knowing the wave function coefficients the current density is
straightforwardly calculated. 

Since the normal reservoir is weakly coupled to the SNS-junction
($d\ll W$), the injected current is small and the current flowing
between the superconductors is only the nonequilibrium Josephson
current. \cite{volkov00} However, the coupling must be large enough so
that the injected quasiparticles do not scatter inelastically before
leaving the junction. Under these assumptions, the distribution of the
injected electrons and holes are governed by the normal reservoir.

The total nonequilibrium Josephson current is naturally parted into
the equilibrium current $I^{eq}$ (at $eV=0$) and the current due to
nonequilibrium, $I^{neq}$.  It has been shown \cite{samuelsson00} that
the nonequilibrium current can be split into two components, one
associated with the nonequilibrium population of the Andreev states
and the other with nonequilibrium mesoscopic fluctuations of the
current. Here, this mesoscopic fluctuation term is neglected, since it
is small compared to term from the nonequilibrium population of the
Andreev levels. The total nonequilibrium current then becomes

\begin{eqnarray}
I\equiv I^{eq}+I^{neq}=\int_{-\infty}^{\infty}dE~i(E)~n_F+\int_{-\Delta}^{\Delta}dE\left[\frac{i(E)}{2}(n^e+n^h-2n_F)\right]
\label{i}
\end{eqnarray}
with $n^{e(h)}=n_{F}(E\mp eV)$ being the distribution functions of
electrons (holes) in the normal reservoir, where
$n_{F}=[1+\exp(E/kT)]^{-1}$.  Clearly, the properties of the current
density $i(E)$ directly determines the nonequilibrium Josephson
current, and the current density will thus be the staring point for
the discussions below.  For energies outside the gap, $E>\Delta$,
$i(E)$ is given by \cite{akkermans91,brouwer97},

\begin{equation}
i(E)=\frac{4e}{h}\frac{d}{d\phi}\mbox{Im}\left(\mbox{tr}~\mbox{ln}\left[1-\alpha^2S(E)r_AS^*(-E)r_A^*\right]\right),
\label{iplusan}
\end{equation}
where $S(E)[S^*(-E)]$ is the electron(hole) scattering matrix for the
normal region and $\alpha=\mbox{exp}[-\mbox{acosh}(E/\Delta)]$ and
$r_A=\mbox{diag}[\mbox{exp}(i\phi/2),\mbox{exp}(-i\phi/2)]$ describe
the Andreev reflection at the NS-interfaces. The dimensions of the
scattering matrices $S(E)$ and $r_A$ are $2N\times 2N$, with
$N=2W/\lambda_F$ the number of transverse transport modes in the
normal region between the superconductors, with $\lambda_F$ the Fermi
wavelength. The current density for energies within the gap is given
by adding a small, positive imaginary part to the energy $E\rightarrow
E+i0$.\cite{doron92,brouwer97} The expression (\ref{iplusan}) then
reduces to the well known result \cite{beenakker91}

\begin{equation}
i(E)=\frac{2e}{\hbar}\sum_m\frac{dE_m}{d\phi}\delta(E-E_m),
\label{ibound}
\end{equation}
where the index $m$ is labeling the bound states given by the equation
$\mbox{det}[1-\alpha^2S(E)r_AS^*(-E)r_A^*]=0$. This form of the
current density is useful when the bound state energy as a function of
phase difference $\phi$ is explicitly known. The current density has
the energy symmetry $i(-E)=-i(E)$.

The junctions studied are in the long limit, $L\gg\xi_0$,~$\xi_0=\hbar
v_F/\Delta$. The nonequilibrium Josephson current in junctions in the
opposite, short limit, has been studied in Ref.\cite{samuelsson00}.
Only classically wide junctions, with many transport modes $N\gg1$, are
considered below. The opposite quantum limit, $N=1$, was studied in
Refs. \cite{vanwees91,chang97,samuelsson97}.

\section{Perfectly transmitting NS-interfaces}
\begin{figure}[h]
\centerline{\psfig{figure=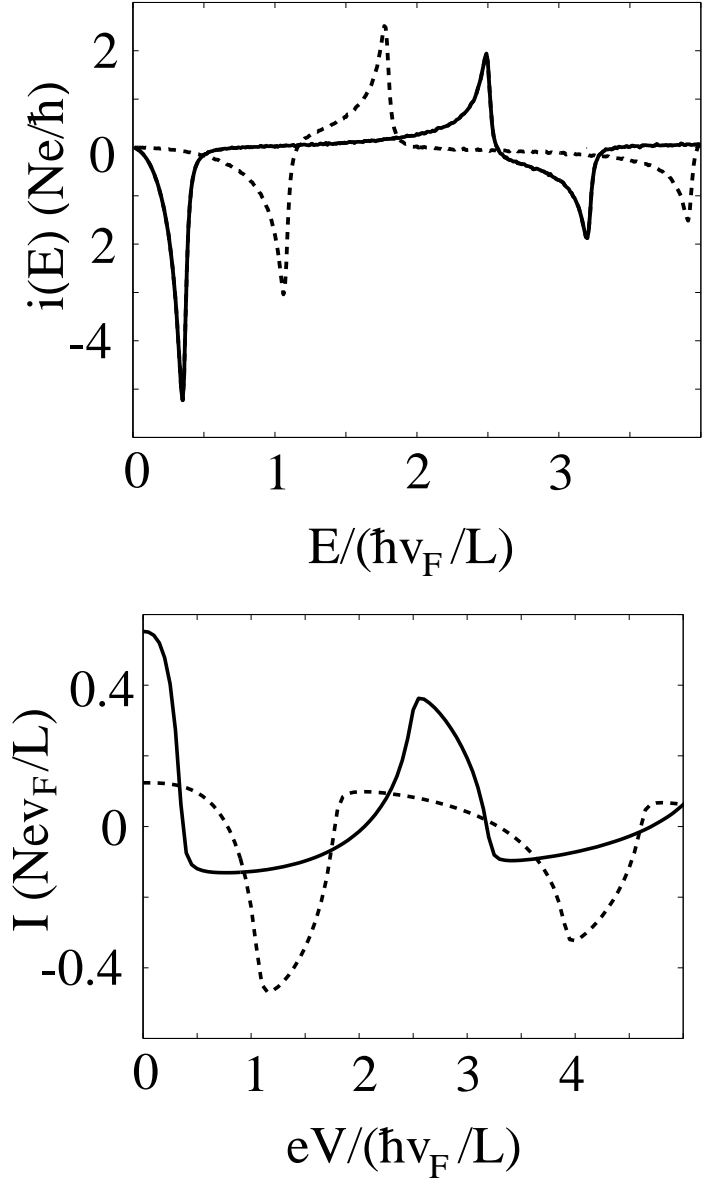,width=0.7\linewidth}}
\caption{Upper: The current density as a function of energy. Lower: The total current as a function of voltage for zero temperature. The phase difference $\phi=\pi/4$ (dashed)and $3\pi/4$ (solid) and $L=10\xi_0$}
\label{longcleandens}
\end{figure}

We first consider the case when there is perfect Andreev reflection at
the NS-interfaces, i.e no normal reflection. For the low energy
levels, $E\ll \Delta$, the bound state energies are given by
\cite{kulik70}

\begin{equation}
  E_{p,n}^{\pm}=\frac{\hbar v_{Fn}}{2L} \left[(2p+1)\pi\pm \phi \right], 
\label{cleanbound}
\end{equation}
where the index $n$ denotes the transverse mode and $p,\pm$ labels the
Andreev levels for a given mode. Due to the hard wall conditions, the
Fermi velocity for each mode $n$ is $v_{Fn}=v_F\sqrt{1-(n/N)^2}$. The
current density for each Andreev level is given by inserting the
expression (\ref{cleanbound}) into (\ref{ibound}). The total current
density is then obtained by first summing over the modes $n$,
equivalent to integrating over angles, and then summing over $p,\pm$,
giving

\begin{equation}
i(E)= N\frac{e}{\hbar}\frac{\hbar v_F}{L}\sum_{p,\pm}\frac{\pm E^{2}\theta(E_{p0}^{\pm}-E)}{(E_{p0}^{\pm})^2\sqrt{(E_{p0}^{\pm})^{2}-E^2}},
\label{ipplus}
\end{equation}
where $\theta$ is the Heavyside stepfunction and $E_{p0}^{\pm}$ is
given from Eq. (\ref{cleanbound}) with $n=0$. The current density for
phase differences $\phi=\pi/4$ and $3\pi/4$ is plotted in Fig.
\ref{longcleandens}. The current density consists of alternating
positive and negative peaks at energies $E_{p0}^{\pm}$. The peaks
arise from the square root singularities of the current density in Eq.
(\ref{ipplus}) and the amplitude of the peaks is decreasing for
increasing energy. 

The expression for the equilibrium part, $I^{eq}$, of the Josephson
current in Eq. (\ref{i}) is known.\cite{kulik70,ishii70,svidzinsky72}
The nonequilibrium part, $I^{neq}$, is straightforwardly calculated
from Eqns. (\ref{ipplus}) and (\ref{i}). It is clear from the form of
the current density in Eq. (\ref{ipplus}) that the total current $I$
will be an oscillating function of voltage, with alternating maxima
and minima, at voltages $eV=E_{p0}^{\pm}$ for zero temperature (see
Fig. \ref{longcleandens}). The period of oscillation is thus $\pi
\hbar v_F/L$. The amplitude of the oscillations, $\Delta
I_p=I(eV=E_{p0}^+)-I(eV=E_{p0}^-)$, in the limit $eV \gg \hbar v_F/L$,
decays with voltage as
\begin{equation}
\Delta I_p \simeq  N\frac{ev_F}{L}\left(\frac{|\phi|\hbar v_F}{LeV}\right)^{3/2}.
\end{equation}
For finite temperatures the amplitude of the voltage oscillations
decreases, and the oscillations are completely washed out for $kT\gg
\hbar v_F/L$.

\begin{figure}[h]
\centerline{\psfig{figure=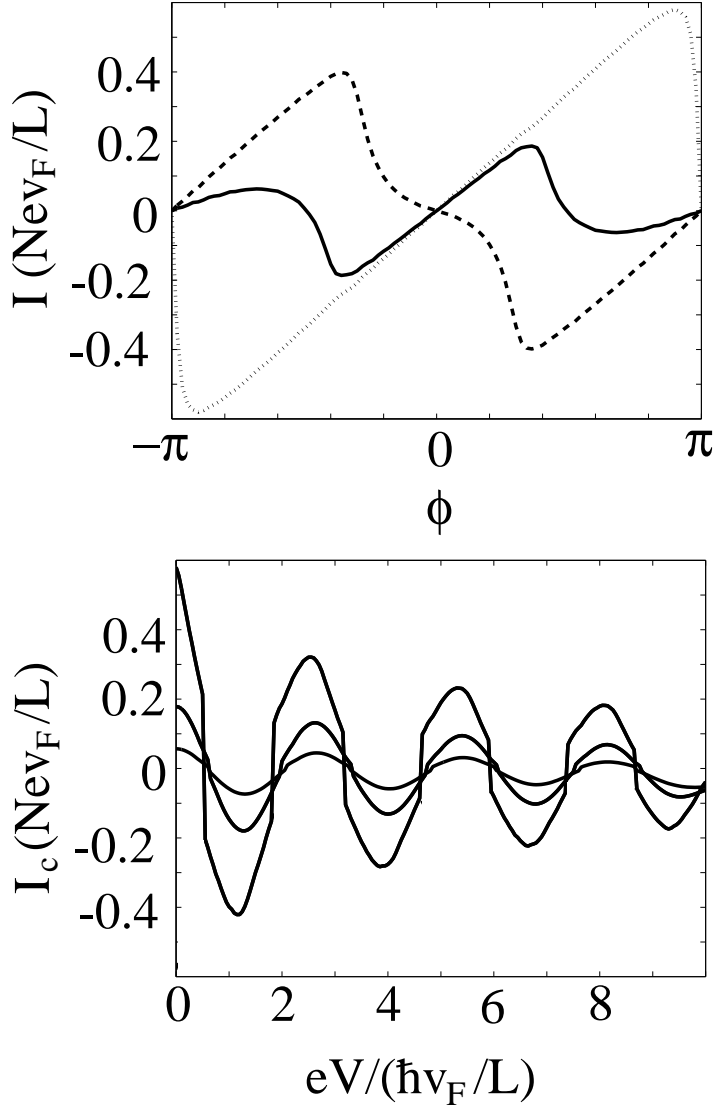,width=0.7\linewidth}}
\caption{Upper: The current phase relationship for different voltages $eV=0$ (dotted), $eV=\hbar v_F/L$ (dashed) and $eV=2\hbar v_F/L$ (solid), temperature $T=0$. Lower: The critical current multiplied by the sign of the critical phase difference $\mbox{sgn}(\phi_c)I_c$ as a function of applied voltage for different temperatures, $kT=0,0.3,0.5\hbar v_F/L$, with decaying amplitude for increasing temperature.}
\label{critclean}
\end{figure}

The equilibrium current $I^{eq}$ is always positive for phase
differences $0<\phi<\pi$ and negative for $-\pi<\phi<0$. In
nonequilibrium, the total current for a given phase difference
$I(\phi)$ changes sign as a function of applied voltage, i.e the
junction becomes a so called $\pi$-junction
\cite{volkov95,baselmans99}. This can bee seen from the current to
phase relationship $I(\phi)$, shown in Fig. \ref{critclean} for
different voltages. It can be noted that for certain voltages, the
current to phase relationship has several local current minima and
maxima.

The critical current is defined as the maximum possible current for
phase difference $-\pi<\phi<\pi$. To study the $\pi$-junction
behavior, the critical current multiplied by the sign of the critical
phase difference $\phi_c$, $\mbox{sgn}(\phi_c)I_c$, is shown in Fig.
\ref{critclean} for different temperatures. 
The critical current multiplied by the critical phase difference
$\mbox{sgn}(\phi_c)I_c$ oscillates between positive and negative
values as a function of voltage, with a period $\pi \hbar v_F/L$, i.e
it shows a typical $\pi$-junction behavior. The amplitude of the
oscillations decreases with increasing voltage. It can be noted that
for zero temperature, there are, for certain voltages, jumps between
the branches of positive and negative critical current, i.e the
critical current $I_c$ never becomes zero. This can be understood from
the current-phase relationship in Fig. \ref{critclean}, where for
certain voltages, the critical phase difference $\phi_c$ jumps between
positive and negative values, changing the sign of
$\mbox{sgn}(\phi_c)I_c$.

\section{Normal reflection at the NS-interfaces}
\begin{figure}[h]
\centerline{\psfig{figure=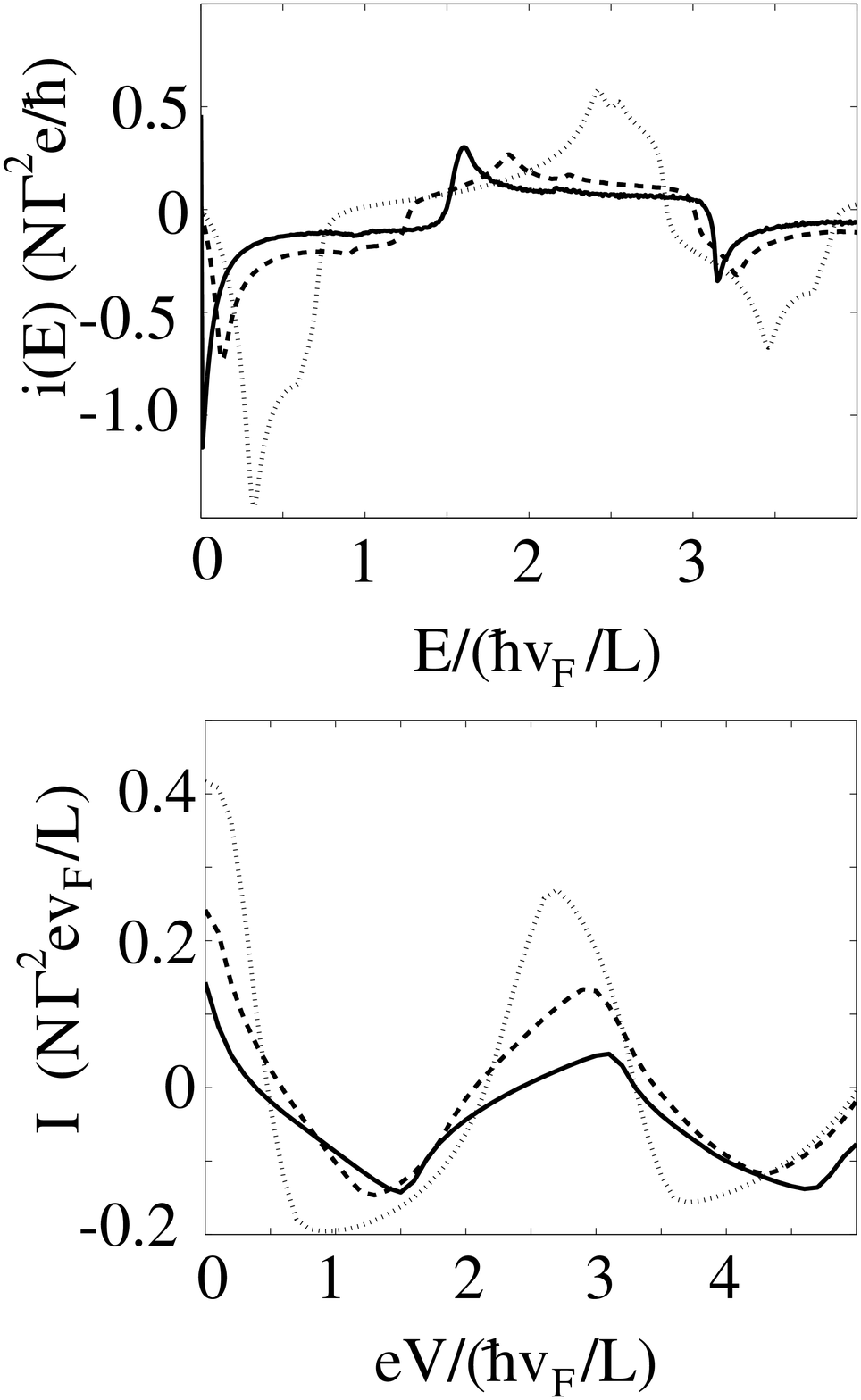,width=0.7\linewidth}}
\caption{Upper: The current density as a function of energy. Lower: The total current as a function of appled voltage. The barrier barrier transparencies $\Gamma=0.1$ (solid), $0.5$ (dashed) and $0.9$ (dotted). The phase difference $\phi=3\pi/4$ and the length $L=10\xi_0$}
\label{barrcurr}
\end{figure}

Normal scattering at the NS-interfaces is taken into account by
introducing effective interface barriers with the transmission
probability $\Gamma$. The low lying bound states energies,
$E\ll\Delta$, are given by\cite{blom98}

\begin{eqnarray}
  E_{p,n}^{\pm}=\frac{\hbar v_{Fn}}{2L} \left[2p\pi \pm \mbox{acos}\left(\frac{4(1-\Gamma)\cos(2k_{Fn}L)-\Gamma^2\cos\phi}{(2-\Gamma)^2}\right) \right].
\label{barrbound}
\end{eqnarray}
The bound state energies oscillate rapidly as a function of length of
the junction, due to the term $\cos(2k_{Fn}L)$ in Eq.
(\ref{barrbound}). The corresponding rapid oscillations of the current
density are averaged out when summing over the transverse
modes.\cite{comment2} The total current density, summed over $n$ and
$p,\pm$, is given by

\begin{eqnarray}
i=\frac{e}{\hbar} \frac{2\Gamma^2 \sin\phi}{\pi} \int_{0}^N dn \frac{\mbox{sgn}[\sin(2EL/\hbar v_{F_n})]}{\sqrt{16(1-\Gamma)^2-[(2-\Gamma)^2\cos(2EL/\hbar v_{F_n})+\Gamma^2\cos\phi]^2}}. 
\end{eqnarray}
The current density as a function of energy is plotted for different barrier transparencies $\Gamma$ in Fig. \ref{barrcurr}. 
The current density has a similar shape, with alternating positive and
negative peaks, as the current density for the junction without
barriers at the NS-interfaces, shown in Fig. \ref{longcleandens}. The
effect of the NS-barriers, apart from the overall decreased current
density amplitude, is that each current density peak is shifted
towards lower energies, as is seen in Fig. \ref{barrcurr}. The current
as a function of voltage, for a fixed phase difference, thus
oscillates with the same period $\pi \hbar v_F/L$ as in the case
without barriers at the NS-interfaces, and with a decreasing amplitude
of the oscillations for increasing voltage (see Fig. \ref{barrcurr}).

\section{Dirty NS-interfaces}

Usually, there is dirt at the NS-interfaces from the junction
processing, leading to diffusive scattering at the interfaces. To
simulate this diffusive scattering, we introduce random scattering
matrices $S_1$ and $S_2$ to model the interfaces. The matrices are
written in the polar decomposition\cite{mello88}

\begin{eqnarray}
S_j=\left( \begin{array}{cc} r_j & t^T_j \\ t_j & r'_j \end{array} \right)=\left( \begin{array}{cc} V_j\sqrt{1-\Gamma_j}V_j^T & V_j\sqrt{\Gamma_j}U_j^T \\ U_j\sqrt{\Gamma_j}V_j^T & -U_j\sqrt{1-\Gamma_j}U_j^T \end{array} \right), 
\label{defs0}
\end{eqnarray}
where the barrier transmittances $\Gamma_j$ are taken to be mode
independent and equal for both barriers. The unitary matrices
$U_j,V_j$ are taken to be independent members of the ensemble of
unitary, symmetric matrices (COE)\cite{beenakker97}. The total
electron scattering matrix for the normal region is given by
\cite{brouwer95}

\begin{equation}
S=r+tS_0(1-r'S_0)^{-1}t^T,~S_0=\left( \begin{array}{cc} 0 & P \\ P & 0 \end{array} \right),
\end{equation} 
where $r=\mbox{diag}(r_1,r_2)$ and similarily for $r'$ and $t$. The
matrix $S_0$ is the scattering matrix for the normal region without
NS-barriers, with the diagonal matrices $P$ with elements
$P_n=\mbox{exp}(i[k_{Fn}+EL/(\hbar v_{Fn})])$, $k_{Fn}$ beeing the
Fermi wave vector. This is inserted in Eq. (\ref{iplusan}) to give the
current density. It can be pointed out that for only forward mode
mixing scattering ($\Gamma=1$), the mode mixing at each interface is
completely reversed by the Andreev reflection, giving the same result
as in the absence of barriers at the NS-interfaces.

\begin{figure}[h]
\centerline{\psfig{figure=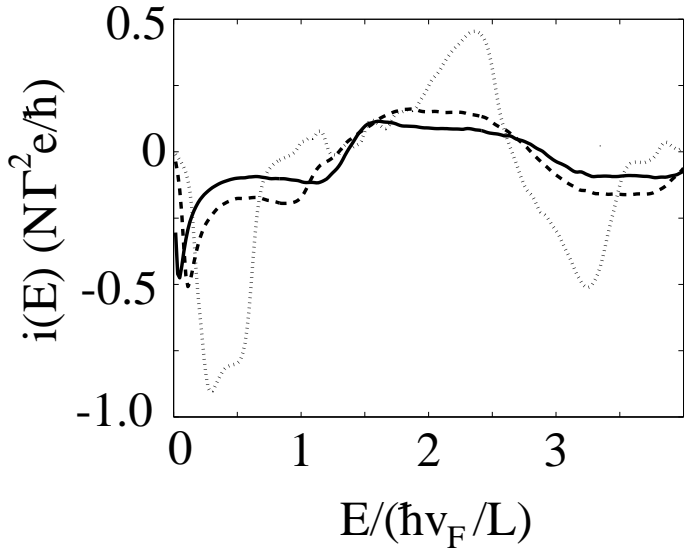,width=0.7\linewidth}}
\caption{The current density as a function of energy for different barrier transparencies $\Gamma=0.1$ (solid), $0.5$(dashed) and $0.9$(dotted). The current density has been calculated by generating $1000$ random matrices of dimension $N=15$. The phase difference $\phi=3\pi/4$ and the length $L=10\xi_0$}
\label{currdirt}
\end{figure}

We are interested in the current density averaged over the random matrices
$U_j,V_j$, which is calculated numerically by generating a large
number of matrices. The current density is plotted in Fig.
\ref{currdirt} for different barrier transparencies $\Gamma$.
  
The main result is that the peak-like structure of the current density
is not substantially changed compared the corresponding specular
barrier case in Fig. \ref{barrcurr}. From this one can draw the
conclusion that the effect of diffusive scattering at the
NS-interfaces does not modify the nonequilibrium Josephson current
qualitatively. The result that the current for a fixed phase
difference oscillates periodically with applied voltage is still
valid. Whether rough 2DEG-sidewalls, giving rise to diffusive boundary
scattering, have a more profound effect on the current, remains to be
investigated. It is known that in the case with the normal region
being a chaotic cavity, the peak-like current density structure is completely
washed out, and a gap opens up in the spectrum.\cite{brouwer97}

\section{Conclusions}
In conclusion, we have studied the nonequilibrium Josephson current in
long two-dimensional ballistic SNS-junctions weakly coupled to a
normal metal reservoir. The total current is given by a convolution of
a single current density $i(E)$ with the quasiparticle distribution
functions [See Eq. (\ref{i})]. Junctions with and without specular
normal scattering at the NS-interfaces and also junctions with
diffusive NS-interfaces are studied. It is found that the current
density in all cases has a peak-like structure, with alternating signs
of the peaks. The resulting nonequilibrium Josephson current for a
given phase difference thus oscillates as a function of applied
voltage, with a period $\pi \hbar v_F/L$, and an amplitude decreasing
with increasing voltage. This behaviour also carries over to the
critical current, which changes sign as a function a voltage, i.e the
junctions displays so called $\pi$-behavior.

\section{Acknowledgements}
This work has been supported by research grants from NFR, TFR, NUTEK (Sweden) and NEDO International Joint Research Grant (Japan).

\end{document}